\documentstyle[preprint,eqsecnum,aps]{revtex}

\begin{document}
\draft \preprint{submitted to Physical Review B}

\title{Hysteresis and Fractional Matching in Thin Nb Films with Rectangular Arrays of Nanoscaled Magnetic Dots}

\author{O. M. Stoll, M. I. Montero}
\address{University of California San Diego, Physics Department 0319, 9500
Gilman Dr., La Jolla, CA 92093, U.S.A.}

\author{J. Guimpel}
\address{Centro At\'omico Bariloche \& Instituto Balseiro,\\
Comisi\'on Nacional de Energ\'{\i}{}a At\'omica \& Universidad
Nacional de Cuyo\\ San Carlos de Bariloche, 8400 R\'{\i}{}o Negro,
Argentina}

\author{Johan J. {\AA}kerman and Ivan K. Schuller}
\address{University of California San Diego, Physics Department 0319, 9500
Gilman Dr., La Jolla, CA 92093, U.S.A.}

\date{\today}
\maketitle
\begin{abstract}

We have investigated the periodic pinning of magnetic flux quanta
in thin Nb films with rectangular arrays of magnetic dots. In this
type of pinning geometry, a change in the periodicity and shape of
the minima in the magnetoresistance occurs for magnetic fields
exceeding a certain threshold value. This has been explained
recently in terms of a reconfiguration transition of the vortex
lattice due to an increasing vortex-vortex interaction with
increasing magnetic field. In this picture the dominating elastic
energy at high fields forces the vortex lattice to form a square
symmetry rather than being commensurate to the rectangular
geometry of the pinning array. In this paper we present a
comparative study of rectangular arrays with Ni-dots, Co-dots and
holes. In the magnetic dot arrays, we found a strong fractional
matching effect up to the second order matching field. In
contrast, no clear fractional matching is seen after the
reconfiguration. Additionally, we discovered the existence of
hysteresis in the magnetoresistance in the crossover between the
low and the high field regime. We found evidence that this effect
is correlated to the reconfiguration phenomenon rather than to the
magnetic state of the dots. The temperature and angular
dependences of the effect have been measured and possible models
are discussed to explain this behavior.

\end{abstract}


\narrowtext

\section{Introduction} \label{sec:level1}

The study of vortex pinning and dynamics in type-II
superconductors is essential for all prospective applications in
which high current densities or magnetic fields are involved. The
rich diversity of different phases found in the mixed state of
high temperature superconductors \cite{Blatter94b,Crabtree97}
shows that vortex pinning and dynamics are also highly interesting
from a fundamental point of view.

Nanolithography provides a method to produce ordered arrays of
artificial pinning centers on the scale of the superconducting
coherence length $\xi$ and the magnetic penetration depth
$\lambda$. With these nanoscaled pinning centers it is possible to
"engineer" the pinning force of a type-II superconductor such that
the critical current $j_c$ is increased for specific magnetic
fields (matching fields). Such arrays can consist of holes
(antidots) \cite{Fiory78,Baert95,Metlushko94,Metlushko99},
magnetic dots \cite{martin97,jaccard98} or of magnetic particles
accumulated in a Bitter decoration experiment \cite{Fasano00}. An
interesting application for periodic antidot arrays is the
reduction of $1/f$-flux noise in SQUIDs \cite{Selders99}. Arrays
of nanoscaled dots have been prepared with various magnetic
\cite{Montero01b,Bael00} and nonmagnetic materials
\cite{Hoffmann00} and with different array geometries such as
triangular \cite{martin97}, square \cite{jaccard98}, Kagome
\cite{Morgan98} and rectangular \cite{martin99b}.

The rectangular arrays seem to be particularly interesting since a
distinct change in the flux-pinning characteristics has been
observed above a certain magnetic threshold field $B_{tr}$
\cite{martin99}. At this field value, the shape of the minima in
the magnetoresistance as well as their periodicity changes. This
behavior has been explained by means of a geometrical
reconfiguration transition of the vortex lattice. In this model,
two competing energies are considered to be important: At low
magnetic field $B$ the pinning energy $E_{pin}$ dominates over the
elastic energy $E_{el}$ of the vortex lattice and the vortices are
dragged onto the artificial pinning centers, by that adjusting to
the underlying rectangular geometry; with increasing field,
$E_{el}$ becomes more important and at the threshold $B_{tr}$ it
forces the vortex lattice back to the intrinsic geometry, which is
assumed to be square.

The pinning mechanism of the vortices by the magnetic dots is
still not completely understood. It is believed that a large
component of the pinning force is of magnetic origin
\cite{Hoffmann00}. Another contribution is likely to come from the
geometrical modulation of the superconducting Nb-film due to the
underlying dots \cite{Montero01b}. Whether the periodic pinning is
mainly caused by a magnetic interaction with the stray fields of
the dots, by the proximity effect or by a combination of different
mechanisms still remains unresolved. Intentional manipulations of
the magnetic domain structure of the dots show that a magnetic
influence does exist in the sense that the pinning force increases
with stronger stray fields \cite{Bael00}.

Another important issue arises, if we consider the fact that a
strong matching effect can be observed in electric transport
measurements with magnetic dots. Since vortices need to move in
order to produce electric dissipation, not only the static
matching but also the dynamics of the flux quanta will play an
important role for the signal. Particularly in the high field part
of the magnetoresistance, the critical current $j_c$ will be low
compared to the applied transport current $j$ so that dynamical
effects can be expected. It is known that a moving vortex lattice
can undergo dynamical phase transitions \cite{Koshelev94} and
order itself at higher flux velocities
\cite{Yaron94,Duarte96,Pardo97}. The systems described in those
references contain a random distribution of defects but the same
effect can be seen also with artificial periodic arrays of pinning
centers \cite{Carneiro00}.

It appears that a study focused on the high field regime of the
rectangular pinning arrays, where the reconfiguration transition
occurs, has the potential of providing insight into the dynamical
nature of the matching effect and the pinning mechanism in
general. Therefore, in this paper we present new experiments done
exclusively on samples with rectangular arrays of magnetic dots.
We focus specifically on the behavior of these samples before and
after the reconfiguration. We find evidence for a fractional
matching effect before the change of regime as well as for new
hysteresis effects occurring in the reconfiguration region. We
discuss these effects in the framework of two possible models: the
geometrical reconfiguration model and a model allowing the dots to
accommodate multiple vortices.

\section{Experiment}\label{sec:level2}

The pinning arrays were prepared by means of e-beam lithography. A
detailed description of the sample preparation can be found in
ref. \cite{martin97,martin98} for the magnetic dots and ref.
\cite{Montero01a} for the holes. In brief, PMMA is spun on top of
a (100) Si substrate. After the e-beam writing process, the
material for the dots is deposited using DC-magnetron sputtering
(Ni) and e-beam evaporation (Co) respectively. Alternatively,
holes can be etched into the substrate using Reactive Ion Etching.
A lift-off process removes the PMMA including the unwanted
material. The remaining dots have a typical thickness of 30 nm and
a typical diameter of 300 nm.  In both cases, the superconducting
Nb film with a thickness of about 100 nm is sputtered on top of
the array.

The results we present in this paper were obtained for three
different samples. For all of these samples we used rectangular
$(a \times b)$ pinning arrays with an aspect ratio $r=b/a = 900 $
nm$/400 $ nm $ = 2.25$. The arrays of samples 1 and 2 were made of
Ni and Co dots respectively, while for sample 3 the pinning array
consisted of 120 nm deep holes in the Si-substrate. The $T_c$ of
the samples was in the range between 6.94 and 8.2 K with a
superconducting transition width of the order of 0.1 K for all
three films. The relevant parameters for the 3 samples are
summarized in Table \ref{table1}. The magnetoresistance was
measured in a standard four probe microbridge geometry with a
bridge width $w = 40$ $\mu$m and a length $L = 50$ $\mu$m between
the voltage leads.

The measurements were performed in a helium cryostat with a 80 kG
superconducting magnet with the magnetic field oriented
perpendicular to the film surface. In some of the measurements a
rotatable sample holder was used to vary the angle $\theta$
between the film normal and the magnetic field. The transport
current was always kept perpendicular to the field direction. It
was applied along the long side $b$ of the rectangular array. Thus
the Lorentz-force always drove the vortices along the short side
$a$.

The voltage drop over the measurement bridge is measured with a
lock-in amplifier which also serves as a supply for the transport
current. The current density is typically in the range from 0.3 to
3 kA/cm$^2$. The electric current for the magnetic field was
provided by a Kepco model BOP20-20M current source and measured as
a voltage drop on a resistor mounted in series with the magnet
leads. With our current experimental setup we can reach magnetic
field resolutions as high as 0.1 G over a total range from -2 kG
to 2 kG. The sweep rate was typically between 0.2 G/s and 2 G/s.
The measurements were found to be independent of the sweep rate
within this range. In order to test for possible effects due to
the AC-transport current, the frequency of the lock-in amplifier
was varied between 17 Hz to 20 kHz, yielding identical results for
the magnetoresistance.

\section{Experimental Results} \label{sec:level3}

\subsection{Fractional Matching} \label{subsec:fractional}

Fig. \ref{MagnetoresNi} shows the positive part of a typical
magnetoresistance curve measured with sample 1. It was recorded
using a magnetic field resolution of 0.1 G which is about a factor
of 50 better than in our previous experiments. In the experiment
shown, the magnetic field was increased from 0 to 600 G with a
rate of about $0.2$ G/s. Clearly two different regimes can be
identified in the curve. At low fields there are sharp and well
defined minima similar to the ones seen in previous measurements
with square \cite{jaccard98,martin97} and rectangular arrays of
magnetic dots \cite{martin99}. The positions can be accurately
described by the $n$th order matching fields $B_n = n
\frac{\phi_0}{a \cdot b}$, where $\phi_0 = 20.7 $ G$\mu$m$^{2}$ is
the magnetic flux quantum and $n$ is an integer number. Thus, from
the experimental values for sample 1 in Fig. \ref{MagnetoresNi},
we can determine $a \times b = 0.421$ $\mu$m$^2$ with an error of
$0.6\%$ resulting from the resolution limit of the electronic
setup.

Apart from the well known integer matching fields, there are
additional structures visible in the low field part of the curve
which is marked with a dashed rectangle. These can be easily
identified as the half integer matching fields for $n =
\frac{1}{2}$ and $n = \frac{3}{2}$. To the best of our knowledge
fractional matching has not been observed yet in rectangular
arrays of magnetic dots. The inset of Fig. \ref{MagnetoresNi}
shows an enlargement of the marked part of the curve. Here, even
finer structures can be observed. The values corresponding to
multiples of the $\frac{1}{4}$ and $\frac{1}{2}$ fractions of the
integer matching fields are highlighted with dashed lines. They
clearly coincide with the respective dips in the
magnetoresistance. The minima for "quarters" are much shallower
than the ones for "halves" in agreement with the fractional
matching seen in hole arrays \cite{Rosseel96}. We emphasize that
fractional matching can be clearly seen up to the 2nd order
matching field. The depth of the minima, and therefore the
corresponding pinning strength, is comparable for the fractional
minima $n=\frac{1}{2}$ and $n = \frac{3}{2}$ (see Fig.
\ref{MagnetoresNi}). In Fig. \ref{pic:hysteresis}, we notice that
the fractional matching effect at half integer fields is also
nicely visible in sample 2, which demonstrates the reproducibility
of this effect independent of the dot material.

At magnetic fields higher than the threshold value $ B_{tr}$, the
behavior changes drastically which has been described by the
previously mentioned reconfiguration transition of the vortex
lattice \cite{martin99}. In this regime, the series of matching
peaks seems to be well described by $B_n = n \frac{\phi_0}{a^2}$,
$a$ being the short side of the rectangle, along which the
Lorentz-force is applied. The vortex lattice literally "loses
memory" of the larger lattice period $b$. Regarding the fractional
matching in this regime, we see that, within our experimental
resolution, there is no observable fine structure. It appears that
the fractional matching is absent or at least much weaker than for
fields $B < B_{tr}$. We discuss possible implications of this
result in section \ref{model} of this paper.

\subsection{Hysteretic Effect} \label{subsec:hysteresis}

We found another interesting effect in our samples with
rectangular arrays of magnetic dots. When the magnetic field is
first increased (or decreased) to a high positive (or negative)
start value and then subsequently swept to zero and further to
negative (or positive) fields, a distinct asymmetry appears in the
magnetoresistance. The result of such an experiment can be seen in
Fig. \ref{pic:hysteresis} (a). For the moment, we concentrate on
the increasing field curve (open squares). Here the recording of
the curve was started at an initial field $B = -150$ G. We observe
a clear asymmetry in the data. It seems that the sharp low field
minimum of the order $n = -2$, which can be expected at around $B
= -100$ G due to symmetry reasons, is missing. On the other hand,
the minimum of the order $n = +2$ at the matching field $B = +100$
G on the positive side is clearly visible. Instead of the minimum
$n = -2$, a much broader peak appears at $B \approx -120$ G which
apparently neither matches the low field periodicity if indexed as
$n = -2$ nor the high field periodicity $\Delta B =
\frac{\phi_0}{a^2}$. It appears that under certain conditions, an
"intermediate state" evolves for magnetic fields close to
$B_{tr}$. If the field is subsequently swept back from $150$ G to
$-150 $ G (filled squares in Fig. \ref{pic:hysteresis} (a)) the
minimum of the order $n = +2$ is missing on the positive side
while the one for $n = -2$ on the negative side is visible.
Apparently, the shape of the magnetoresistance curves depends
strongly on the magnetic history of the sample. It is important to
point out, that in the experiment in Fig. \ref{pic:hysteresis}
(a), the field has been increased above the threshold value
$B_{tr}$ and decreased below $-B_{tr}$, respectively. The fact
that the observed hysteresis appears at magnetic fields close to
the threshold values $\pm B_{tr}$ suggests, that it may have to do
with the reconfiguration transition.

In order to confirm this conjecture, we repeated the experiment
keeping the magnetic field in the range between the threshold
values $B_{tr} \approx \pm 140$ G, obtaining the results shown in
Fig. \ref{pic:hysteresis} (b). In this case, contrary to the data
in Fig. \ref{pic:hysteresis} (a), we obtain a fully reversible
magnetoresistance curve except for a small deviation on the
positive side which is probably due to a small temperature drift
during the measurement. This proves that the hysteresis is related
to the reconfiguration phenomenon.

Hysteretic effects in conjunction with periodic pinning phenomena,
have also been reported in literature. Therefore possible
implications for our results have to be discussed. A potential
microscopic origin of these effects is the hysteresis due to the
alignment of the magnetic moment of the dots, when the external
magnetic field is swept beyond the coercive field for the
perpendicular direction. An asymmetry of the critical current for
samples with arrays of magnetic dots due to this mechanism has
been reported in ref. \cite{Morgan98}. However, in those
experiments the dots were much thicker (110 nm) and had a smaller
diameter (120 nm). Thus, the shape anisotropy can be expected to
be much smaller than for the geometry in our experiments ($d_{dot}
= 300$ nm, $t_{dot} = 30 - 40$ nm, see Tab. \ref{table1}). Still,
the magnetic field used in ref. \cite{Morgan98} to magnetize the
dots perpendicular to the film surface was around 3.5 kG, whereas
in our experiment, Fig. \ref{pic:hysteresis} (a) and (b), it was
always kept below 150 G. Because of the larger shape anisotropy,
we can expect our dots to have their entire magnetic moment
in-plane (parallel to the film surface). Also, a pronounced
asymmetry effect in the magnetization of Pb-films with square
arrays of Pt/Co/Pt dots, which have their magnetic moment
perpendicular to the plane, has been observed
 \cite{Bael00}. In this work no asymmetry in the magnetization vs. B characteristics
 was visible for dots with a magnetic moment
 in-plane. However, there was a small difference in
 the behavior before and after an initial magnetization procedure.
 This has been shown to be due to the formation of single domain
 configurations out of the as-grown multidomain arrangement.
AFM-imaging of domain walls on magnetic dots similar to the ones
in our samples makes the existence of such domains likely
\cite{Wittborn00}. Since the hysteresis in our case is
reproducible from measurement to measurement and since we don't
find any difference between the initial sweep and the consecutive
experiments, we conclude that a domain switching process either
doesn't occur or has no visible influence on our results.

To completely exclude an effect due to a change in the effective
magnetic moment of the dots, we repeated the above described
experiment with a sample consisting of an array of holes in the
substrate (sample 3). If the effect would be due to the
magnetization of the dots, this sample should obviously not show
the hysteretic magnetoresistance. From the plot in Fig.
\ref{pic:hysteresis} (c) it becomes clear that the hysteresis is
also present and, consequently, that the magnetic moment of the
dots does not play a role in its origin. The field sweeps (open
squares from $-200$ to $200$ G, filled squares from $200$ to
$-200$ G) show hysteresis like the ones with the magnetic dot
arrays (samples 1 and 2). Here, probably due to the less effective
pinning of the holes compared to the magnetic dots, the
reconfiguration transition occurs already after the first order
matching field ($n = 1$). This behavior has already been described
in detail elsewhere \cite{Montero01b}.

The results obtained so far suggest that the hysteretic effect is
correlated to the reconfiguration transition appearing in
rectangular arrays of magnetic dots. The fact that the
magnetoresistance and the critical current for square arrays,
where no reconfiguration is expected, have been found to be
symmetric without showing hysteresis within the available
experimental resolution \cite{martin97,Hoffmann00} is in agreement
with this result.

\subsection{Angular and Temperature Dependence} \label{subsec:AngTemp}

The dependence of the periodic pinning on the angle $\theta$
between film normal and magnetic field has been studied for square
arrays of magnetic dots \cite{martin97}. It was found that only
the component of the field perpendicular to the film surface
$B_{\perp}$ matters for the vortex system \cite{martin97}, i. e.
that the applied magnetic field $B$ is effectively reduced by a
factor $\cos \theta$. However, this has not yet been confirmed for
rectangular arrays of magnetic dots. In this geometrical
configuration it is especially interesting to study, whether the
reconfiguration transition and/or the high field behavior also
depend only on the normal component of the field. The geometry of
our experiment on rectangular arrays is sketched in Fig.
\ref{pic:angular} (lower inset). The magnetic field was tilted
along the short side $a$ of the rectangle. Therefore the
Lorentz-Force, resulting from the current applied along the long
side $b$, remains always parallel to the short side $a$. Fig.
\ref{pic:angular} shows a series of magnetoresistance curves of
sample 2 for values of $\theta$ between $3^{\circ}$ and
$72^{\circ}$ as a function of the perpendicular component of the
applied field $B_{\perp} = B \cos \theta$. We note again a
pronounced asymmetry in the curve, as described earlier in this
section. Furthermore the positions of the peaks scale nicely with
the $1/\cos \theta$ up to high orders (see upper inset of Fig.
\ref{pic:angular}). This behavior is identical to that of the
square arrays of magnetic dots. Moreover, the position of the
threshold field $\pm B_{tr}$ does not depend on the angle
$\theta$. This means that both the reconfiguration transition and
the behavior of the vortex system after the reconfiguration only
depend on $B_{\perp}$. Apparently, for the peak position only the
number of vortices per unit cell of the periodic array is
important, similar to what has already been found for the square
geometry. Nevertheless, there is a notable difference in the {\it
absolute} value of the magnetoresistance if we compare the parts
of the curve above and below $|B_{tr}|$. The low field part $|B| <
B_{tr}$ is very stable and reproducible when scaled with $B \cos
\theta$. In contrast, in the high field section the resistance
increases considerably with increasing $\theta$. It is striking,
that the stability of the low field regime stretches out to about
the same field values on the positive and negative side of the
x-axis regardless of the fact that there is a minimum missing on
the positive side. Up to now, we don't have a conclusive
explanation for this behavior, although it could be the beginning
of the transition to the normal state due to the fact that the
total applied field $B$ comes close to the $B_{c2}$ value of our
film. This explanation is slightly contradictory with the fact
that the properties scale as the normal component of the field. It
could indicate that the film thickness is less than but not
negligible when compared to $\lambda$.

The temperature dependence of the asymmetry and of the sudden
change in the periodicity of the magnetoresistance minima for the
rectangular arrays can give important clues about the mechanisms
involved in causing these effects. Therefore, we recorded a series
of curves at different temperatures $T$ close to $T_c$, which are
shown in Fig. \ref{pic:temperature}. For clarity, the curves are
shifted with respect to each other along the voltage-axis. The
position of zero magnetic field $B = 0$ is marked with a solid
line. Because of the change of the critical current $j_c$ with
temperature, the transport current for each measurement has been
adjusted such, that the dissipation level at a given magnetic
field remained the same for all curves. For the experiment shown
in Fig. \ref{pic:temperature} we used a voltage criterion of $10$
$\mu$V at a field of $800$ G. For all curves, the magnetic field
has first been increased to a positive start value and
subsequently swept to negative values. Once again, we can see the
typical asymmetry described in section \ref{subsec:hysteresis}.
For all of the curves, the minimum for the positive second order
matching field $n = +2$ is missing. We also observe a different
temperature dependence of the positive and negative sections of
the curves. On the positive side, the minimum with $n = +2$, which
is suppressed for $T = 7.701$ K, starts to develop with decreasing
temperature until it is clearly visible for the lowest temperature
$T = 7.393$ K. Because of the asymmetry the change on the negative
side affects the minimum with $n = -3$. Here, the minimum seems to
become more pronounced (deeper) with decreasing temperature as
well. However, the change is less dramatic than for $n = +2$. If
we take a look on the overall shape of the curve, the low field
part $|B| < B_{tr}$ varies little with temperature on both sides.
In contrast, the part after the reconfiguration $|B|
> B_{tr}$ seems to be strongly $T$-dependent. The minima are
effectively "washed out" with decreasing $T$. This indicates, that
random pinning gains importance compared to the artificial
periodic pinning as the temperature is further lowered below
$T_c$. Apparently the random defects are much more important for
the behavior after the reconfiguration transition than before.
This is a strong indication, that different mechanisms are
responsible for the matching phenomenon in these two parts of the
curve.

\section{Model Discussion} \label{model}

In the discussion of our results we will again distinguish between
the two qualitatively different parts visible in our data. In the
low field regime, we have a well defined series of resistance
minima with a periodicity related to the dot unit cell area. This
part will be discussed in section \ref{subsec:LowField}.

The high field minima are less well defined and their periodicity
appears to be exclusively related to the side of the dot unit
cell, along which the Lorentz-force is applied. The transition
between the two phases shows a characteristic hysteretic behavior.
The high field part and the transition between the two regimes,
will be discussed in section \ref{subsec:HighField}.

\subsection{Low Field Phase}\label{subsec:LowField}

The low field data is usually described in terms of two possible
models. The first one, which we label as the ``matched lattice''
model \cite{martin98}, assumes that only one vortex can be pinned
to a magnetic dot. The vortex lattice matches the dot array and
the excess vortices are forced into interstitial symmetry
positions of the underlying array. Here, the magnetoresistance
minima are directly equivalent to maxima in the critical current.
These are due to the fact that at integer numbers of vortices per
unit cell of the dot lattice, there are no free interstitial
positions for the vortices to jump to. In the second model, which
we label as the ``multivortex model'' \cite{Baert95}, each
magnetic dot is able to accommodate more than one vortex. This can
happen either in the form of multiple confined vortices or as a
single multiquanta vortex. In this model the maxima in the
critical current are understood in similar terms. Now the vortices
jump between the magnetic dots and an increase in the critical
current occurs whenever the number of vortices is the same on each
dot, i.e. again at integer number of vortices per unit dot
lattice. Pinning of multiple vortices to a single pinning center
is possible, if the saturation number $n_s$ is larger than 1. For
an isolated hole in the superconductor, it can be estimated using
the expression $n_s \approx \frac{\kappa r}{2 \lambda}$
\cite{Mkrtchyan72}. Here, $\kappa$ is the Ginzburg-Landau
parameter and $r$ is the radius of the pinning center. In a
periodic array however, $n_s$ can be expected to be higher because
the interaction with the next neighbor vortices in the lattice is
not negligible. Therefore, the saturation number will also depend
on the geometry of the pinning array.

For both models, the origin of the fractional matching peaks has
an explanation similar to the one sketched above for the integer
order matching peaks. A symmetrical periodic vortex structure is
formed and, in order to move the vortices, this symmetrical
structure has to be broken. However, the periodicity of this
fractional order structure is larger than one dot lattice unit
cell, and thus the critical current enhancement (or resistivity
reduction respectively) is smaller.

The experimental data obtained at low fields seems to favor the
multivortex model. A schematic illustration of the two models is
shown in Fig. \ref{pic:model}. For the matched lattice model one
expects the vortices to be pinned more strongly for magnetic
fields below the first order minimum than above. This is due to
the fact, that the vortices are interacting directly with the dots
for $B < B_1$ but reside in interstitial potential wells for $B >
B_1$ (see Fig. \ref{pic:model} (a) for $n =1$ and $n =2$).
Consequently, the resistivity should show a substantial increase
immediately above the first order matching field
\cite{Reichhardt01a}, contrary to the experimental data.

In contrast, the multivortex model should show basically a field
independent pinning, since the vortices are confined to the
magnetic dots \cite{Reichhardt01a} (see Fig. \ref{pic:model} (b)
for $n = 1$ and $n = 2$). This means that the strength of the
fractional matching for $B > B_1$ is expected to be comparable to
the one for $B < B_1$. In reality it is slightly weaker due to the
additional repulsive interaction with the other flux quantum which
is allocated to the dot. This picture is in agreement with the
experimental data for $B < B_{tr}$.


\subsection{Transition Region and High Field Phase}\label{subsec:HighField}

The transition between the two regimes, which up to now we have
called a "reconfiguration transition", has a different explanation
in the two models. For the multivortex model a change in behavior
can be expected when the saturation number $n_s \approx 2$ for the
dots is reached \cite{Bezryadin96,Mkrtchyan72}. Above this
saturation field, additional vortices have to sit at interstitial
positions as illustrated in Fig. \ref{pic:model} (b) ($n = 3$).
Therefore, in this scenario, the transition is an indication of
the formation of interstitial vortices. Such a coexistence between
multiquanta and interstitial vortices has been seen in Bitter
decoration experiments with arrays of holes \cite{Bezryadin96}.
The absence of a fractional matching effect above $B_{tr}$, which
has been described in Section \ref{subsec:fractional}, could be
another indication for a much weaker pinning due to these
interstitial vortices. It has been shown in theoretical
simulations, that these interstitials tend to move in channels and
produce no or much weaker fractional matching peaks
\cite{Reichhardt01a,Reichhardt00b}. However, as already discussed
above for the matched lattice model, the critical current of the
interstitial vortices should be lower than that of the ones pinned
at the dots, and consequently a substantial increase of
resistivity should be observed immediately above the change of
regime. In contrast, our data in Fig. \ref{MagnetoresNi} and Fig.
\ref{pic:temperature} shows that the absolute resistance value
stays about the same or even decreases after the value $B_{tr}$ is
exceeded.

The presence of hysteresis in the change of regime implies the
existence of an energy barrier between the two configurations and
thus may involve a first order transition. For the multivortex
model it has been pointed out \cite{Bezryadin96} that the
transition between multiple vortices and interstitial vortices is
indeed a first order phase transition, which could explain the
hysteretic magnetoresistance.

In the matched lattice model, it has been proposed \cite{Martin00}
that the change of regime occurs as soon as the elastic energy of
the vortex lattice $E_{el}$ dominates over the pinning energy
$E_{pin}$. Then the vortex lattice reconfigures from a
commensurate rectangular to a square configuration as shown in
Fig. \ref{pic:model} (a) (n = 3). This suggestion is based on the
experimental fact that the series of maxima observed at high
fields show a periodicity, which seems to be related the side of
the rectangle along which the force is applied \cite{Martin00}. In
this picture, the minima signal the matching of the vortex lattice
parameter to the dot lattice parameter along the movement
direction, $a = n\;a_{o} $ leading to a \textit{dynamical}
pinning. However, given that the vortex lattice parameter scales
as $B^{ - \raise.5ex\hbox{$\scriptstyle 1$}\kern-.1em/
\kern-.15em\lower.25ex\hbox{$\scriptstyle 2$}} $ , $a_{o} = \sqrt
{{\frac{{\phi _{o}} }{{B}}}} $, this line of reasoning would give
rise to a \textit{quadratic} rather than a \textit{equally spaced}
series of peaks, $B_{n} = n^{2}{\frac{{\phi _{o}} }{{a^{2}}}}$.
This prediction is obviously very different from the
\textit{linearly equally spaced} series of peaks found
experimentally \cite{Martin00}. This argument leads to the
conclusion that the explanation of these features cannot be
achieved with a simple lattice matching argument. If this idea is
to be further developed, other mechanisms such as a reorientation
of the vortex lattice for different fields or a loss of coherence
due to a mismatch along the applied current direction with a
simultaneous matching along the force direction have to be
considered.

The hysteresis cannot be easily understood in the matched lattice
model. To pin and depin the lattice, the vorticity at the dots has
to change between 1 and 0, but this change does not present an
energy barrier for a hole in a superconductor \cite{Bezryadin95}.
If the observed change of regime is indeed an indication for a
change of the lattice geometry, then a barrier should exist
between these two configurations.

Up to this point we have discussed the results in terms of
\textit{static} models, implicitly thinking of a vortex lattice
geometrically commensurate or incommensurate with the fixed dot
array. However, it has already been mentioned that there is
clear-cut evidence pointing to the presence of \textit{dynamical}
effects in the magnetoresistance. For example, it has been
recently shown that the observed features in the magnetoresistance
are strongly dependent on the vortex velocity \cite{Velez01}. Also
it has been found that the position of the minima for samples with
rectangular arrays of magnetic dots depends on the direction of
the applied current \cite{martin98}. The importance of dynamical
ordering of the vortex lattice under the influence of periodic
pinning has been recently stressed in studies of the vortex
lattice structure using Bitter pinning \cite{Fasano00}. It was
found that if a vortex lattice is driven by a change in the
direction of the applied field, the very weak periodic pinning
caused by a pattern of Fe clumps produced in a first Bitter
decoration experiment can dominate over the bulk pinning in spite
of being orders of magnitude smaller.

Recent simulations of driven vortex movement in the presence of
rectangular arrays of pinning centers
\cite{Reichhardt00a,Reichhardt00b} clearly show the formation of
channels between the rows of dots. These results offer an
interesting and intriguing possibility for the analysis of our
experimental results. In this new scenario, channels of moving
vortices between two consecutive rows of dots would dynamically
order to form a lattice, infinite in the direction of movement but
finite in the perpendicular direction.  In this situation, the
vortices pinned to the dots form a repulsive periodic potential.
Thus, the edges of the moving lattice experience a periodic
perturbation the timescale of which depends only on the lattice
parameter of the array along the movement direction. This
corresponds to a frequency $f = a / v$, where $v$ is the lattice
velocity. A similar effect has been seen for the periodic pinning
in superlattices in which the vortices are moving perpendicular to
the layers \cite{Gurevich96}. However, at this moment it is not
clear to us how this perturbation and its interaction with the
dynamical states of the moving lattice would translate to the
structure observed for the high field phase.

\section{Summary and Conclusions} \label{summary}

We have investigated the periodic vortex pinning in rectangular
arrays of magnetic dots. In our magnetoresistance measurements we
found a strong fractional matching effect up to the second order
matching field. For magnetic fields larger than a threshold value
$B_{tr}$ a distinct change in behavior occurs in this type of
pinning array. Above this field, the fractional matching is absent
or at least much weaker than below $B_{tr}$.

We also observed an interesting hysteretic effect in the
magnetoresistance curves when the magnetic field is swept above
the reconfiguration threshold and back again. We showed that this
effect does not appear if this threshold $B_{tr}$ is not exceeded.
Additionally, we find the same effect also in samples with
nonmagnetic pinning centers. Therefore, we conclude that it is
correlated to the "transition" $B_{tr}$ rather than to the
magnetic moments of the dots.

Our experimental data suggests that a model including multivortex
pinning explains our data better than a model based on the
formation of interstitials. We argue that the observed transition
could be due to a crossover from multivortex to interstitial
vortex pinning. This explanation would explain the observed
hysteretic behavior in terms of a first order transition. It is
likely that also dynamical effects such as dynamic ordering of the
lattice and the formation of channels have to be taken into
account in order to explain the high field behavior of the
magnetoresistance. In order to resolve these issues, experiments
which directly image the vortex configuration and correlations
with the transport measurements could be useful.

\acknowledgments

This work was supported by DOE and NSF. OMS acknowledges support
by the Deutscher Akademischer Austauschdienst (DAAD). JG
acknowledges support from the Fulbright Commission and Fundaci\'on
Antorchas in the form of a Fulbright fellowship during his visit
to UCSD. JG is also a CONICET (Argentina) fellow. MIM thanks the
Spanish Secretar\'{\i}a de Estado de Universidades e
Investigaci\'on for supporting her stay at UCSD.

\bibliographystyle{prsty}

\bibliography{bibliographie}



\begin{figure}
\caption{Magnetoresistance of sample 1 (Ni-dots) measured at $T =
7.8$ K with $I = 0.2$ mA. Shown is only the positive part of the
curve. The dashed rectangle marks the low field part. The inset
shows a magnification of the part of the curve marked with the
dashed rectangle. The magnetic field is normalized with the first
order matching field $B_1$. The field values corresponding to
fractional matching are highlighted with dashed lines.}
\label{MagnetoresNi}
\end{figure}

\begin{figure}
\caption{Hysteretic Effect of the magnetoresistance. The curves
with the open squares and the filled squares correspond to a
increase and a decrease of the magnetic field, respectively. (a)
Magnetoresistance of sample 2 (Co-dots) measured at $T = 7.8$ K
with $I = 0.2$ mA. In this experiment the field initially exceeded
the values $\pm B_{tr}$ for the respective curves before the
recording was started. (b) Magnetoresistance for the same sample
at $T = 7.8$ K and $I = 0.2$ mA. Here the field was kept below
$\pm B_{tr}$. (c) Magnetoresistance of sample 3 consisting of an
array of holes measured at $T = 6.5$ K with $I = 0.2$ mA.
\label{pic:hysteresis}}
\end{figure}

\begin{figure}
\caption{Angular dependence of the magnetoresistance for sample 2.
The x-axis is normalized with the projection of the field on the
normal to the film surface $B \cos \theta$. The upper inset shows
a plot of the position $B$ of the minima vs. $1/cos \theta$ for
the order $n = 1$ (filled triangles) through $n = 6$ (open
circles). The lower inset shows a sketch of the geometry used for
the angular dependent experiments. Here, $\vec{n}$ is the normal
to the sample surface, $a$ and $b$ are the short and the long side
of the rectangle. $B_{\perp}$ corresponds to the component of the
magnetic field perpendicular to the surface and $B$ is the total
applied field. \label{pic:angular}}
\end{figure}

\begin{figure}
\caption{Temperature dependence of the magnetoresistance for
sample 2. For each curve the current was adjusted to give the same
voltage of $10$ $\mu$V at a field of $800$ G. For clarity, the
curves are shifted with respect to each other along the voltage
axis. \label{pic:temperature}}
\end{figure}

\begin{figure}
\caption{Comparison of (a) the "matched lattice" model and (b) the
"multivortex" model. The sketch shows the situation schematically
for the matching fields of orders $n = 1$ and $n = 2$ before and
for the order $n = 3$ after the reconfiguration. The transition is
symbolized by a dashed line. This situation resembles the one
found in our experiments. \label{pic:model}}
\end{figure}

\begin{table}
\caption{Sample characteristics. $d_{dot}$ is the dot diameter,
$t_{Nb}$ and $t_{dot}$ are the Nb-film thickness and the dot
thickness respectively. In the case of the holes $t_{dot}$ is the
hole depth. \label{table1}}

\begin{tabular}{lcccccc}

& Dot Material & $T_c$ & $\Delta T_c$ & $d_{dot}$  & $t_{Nb}$ &
$t_{dot}$\\ \tableline

Sample 1 & Ni & 8.2 K & 0.093 K & 300 nm & 75 nm & 38 nm \\

Sample 2 & Co & 8.3 K & 0.115 K & 300 nm &75 nm & 30 nm \\

Sample 3 &  holes & 6.94 K& 0.12 K& 300 nm  & 80 nm  & 120 nm

\end{tabular}
\end{table}

\end{document}